\documentclass[11pt,superscriptaddress,preprint,tightenlines,floatfix]{revtex4}
\pdfoutput=1 
\usepackage{graphicx}

\begin{document}

\title{Digital control of force microscope cantilevers using \\
       a field programmable gate array}


\author{Jonathan P. Jacky}
\email{jon@washington.edu}
\altaffiliation[Mailing address: ]{University of Washington, 
Mechanical Engineering, Box 352600, Seattle, Washington, 98195-2600}
\affiliation{Department of Orthopaedics and Sports Medicine, 
University of Washington}
\author{Joseph L. Garbini}
\affiliation{Department of Mechanical Engineering, University of Washington}
\author{Matthew Ettus}
\affiliation{Ettus Research LLC}
\author{John A. Sidles}
\affiliation{Department of Orthopaedics and Sports Medicine,
University of Washington}

\begin{abstract}
This report describes a cantilever controller for magnetic resonance
force microscopy (MRFM) based on a field programmable gate array
(FPGA), along with the hardware and software used to integrate the
controller into an experiment.  The controller is assembled from a
low-cost commercially available software defined radio (SDR) device
and libraries of open-source software.  The controller includes a
digital filter comprising two cascaded second-order sections
("biquads"), which together can implement transfer functions for
optimal cantilever controllers.  An appendix in this report shows how
to calculate filter coefficients for an optimal controller from
measured cantilever characteristics.  The controller also includes an
input multiplexer and adder used in calibration protocols.  Filter
coefficients and multiplexer settings can be set and adjusted by
control software while an experiment is running.  The input is sampled
at 64 MHz; the sampling frequency in the filters can be divided down
under software control to achieve a good match with filter
characterisics.  Data reported here were sampled at 500 kHz, chosen
for acoustic cantilevers with resonant frequencies near 8 kHz.  Inputs
are digitized with 12 bits resolution, outputs with 14 bits.  The
experiment software is organized as a client and server to make it
easy to adapt the controller to different experiments.  The server
encapusulates the details of controller hardware organization,
connection technology, filter architecture, and number representation.
The same server could be used in any experiment, while a different
client encodes the particulars of each experiment.

\end{abstract}

\maketitle

\section{Introduction}

This report describes a cantilever controller for magnetic resonance
force microscopy (MRFM) based on a field programmable gate array
(FPGA).  The controller is assembled from a low-cost commercially
available software defined radio (SDR) device, libraries of
open-source software, and some additional software that we have
written.  In addition to the controller itself, we also describe
software that integrates the controller into an experiment.  We 
show how to calculate filter coefficients for an optimal controller
from measured cantilever characteristics.  Finally, we present data
comparing calculated and measured controller performance.

\begin{figure}
\begin{center} 
\includegraphics[width=5in]{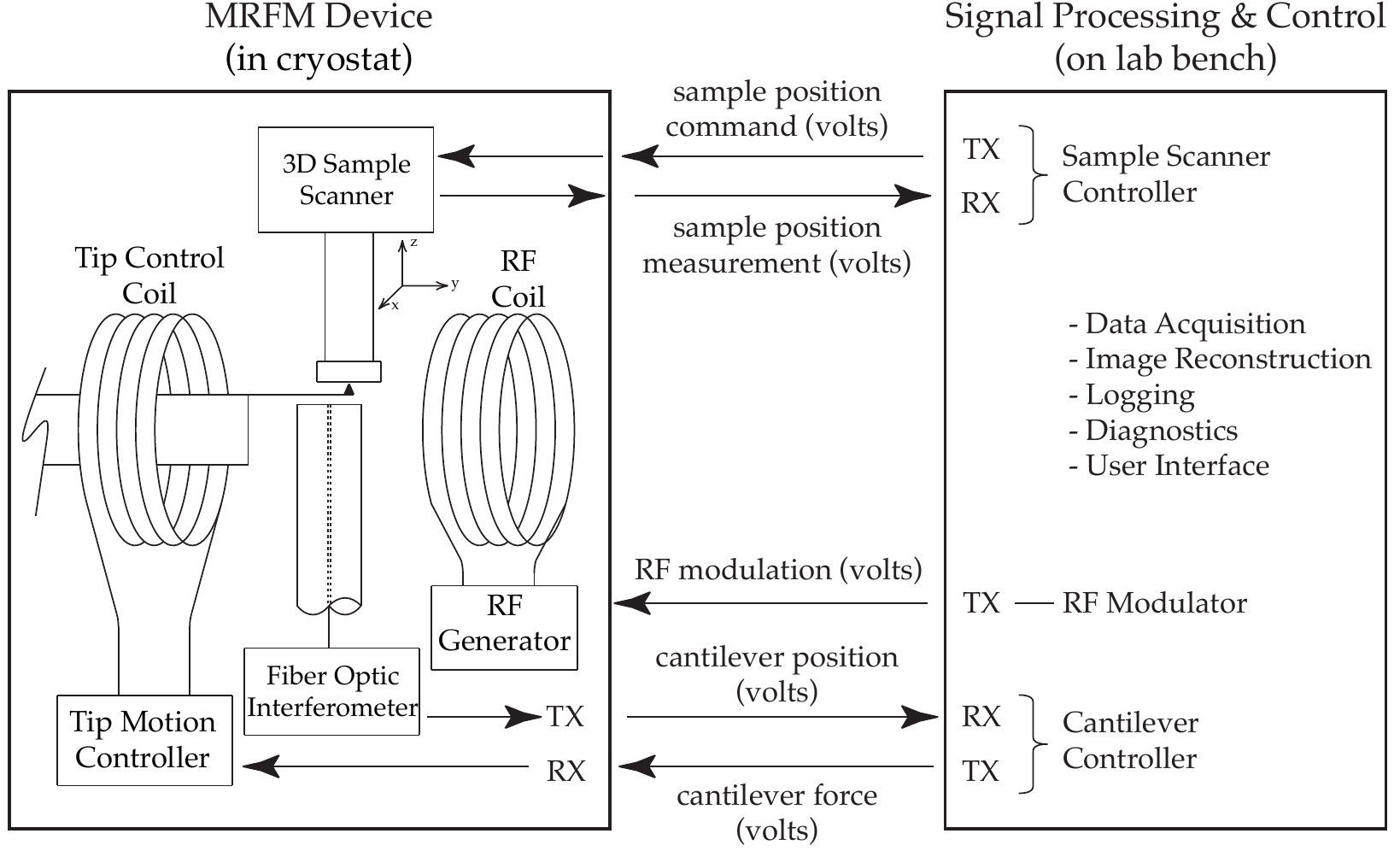} 
\caption{MRFM experiment\label{fig:mrfm-block}}
\end{center}
\end{figure}

Force microscope cantilevers are mechanical oscillators that require
feedback control.  In MRFM, 
cantilevers sense the magnetic force exerted by spins in the
sample~\cite{Sidles:91,Sidles:95}. Cantilevers for detecting such
small forces must have low spring constants and high quality factors,
but these result in large deflections, long ring-down times, and
narrow bandwith, which hinder image acquisition.  These effects can be
mitigated by feedback control without affecting the signal-to-noise
ratio for force measurement~\cite{Garbini:96}.  In the apparatus shown
in Fig.~\ref{fig:mrfm-block}, the cantilever controller would be
connected between RX and TX in the lower right corner (RX is the
controller input).

Cantilever controller design is discussed
in~\cite{Garbini:96,Bruland:96,Bruland:98a,Degen:06}; details of our
design appear in the appendix.  Early controllers
used analog circuits including operational
amplifiers~\cite{Bruland:96,Bruland:98a}; recent controllers use
specialized microprocessors called Digital Signal Processors
(DSPs)~\cite{Chao:04,Degen:06}.  In this report we describe a new
digital cantilever controller based on an FPGA.

An FPGA contains an array of identical digital logic and memory
elements which can be connected together under software control to
form, in effect, a custom integrated circuit.  FPGA-based signal
processors can have several advantages over microprocessors (including
DSPs).  Computations can be highly parallel (instead of executing a
sequential instruction stream); functions executing in parallel do not
affect each other's speed, so there can be greater throughput
even at lower clock speeds.  Latencies can be lower because each input
and output can be connected directly to the FPGA (instead of
connecting them all through a shared bus).  Precision (word length)
can be tailored to computational requirements (instead of relying on a
few fixed word sizes).

This controller accommodates input and output signals up to 2 volts
peak-to-peak (-1V to +1V), digitized with a resolution of 12 bits
(input) and 14 bits (output).  The sampling frequency can be 64 MHz,
or can be {\em decimated} (divided down) under software control.  The
controller demonstration described here samples at 500 kHz, chosen for
acoustic cantilevers with resonant frequencies near 8 kHz.  The
controller includes an infinite impulse response (IIR) digital filter
comprising two cascaded second-order sections (biquadratic sections,
commonly called ``biquads'').  The signal and the filter coefficients
are represented by 24-bit integers.  The experiment software
translates floating point to appropriately scaled 24-bit integers.

In addition to the filter, the controller also includes an input
multiplexer and adder that selects and optionally adds two input
signals in all combinations.  This is helpful for some
calibration protocols.

The controller characteristics (filter coefficients, multiplexer
settings etc.) can be set and adjusted by control software while an
experiment is running.

We expect that this controller will acccommodate future
microscopes in our laboratory with modest effort, and we believe it
could be used elsewhere as well.

\section{Hardware}

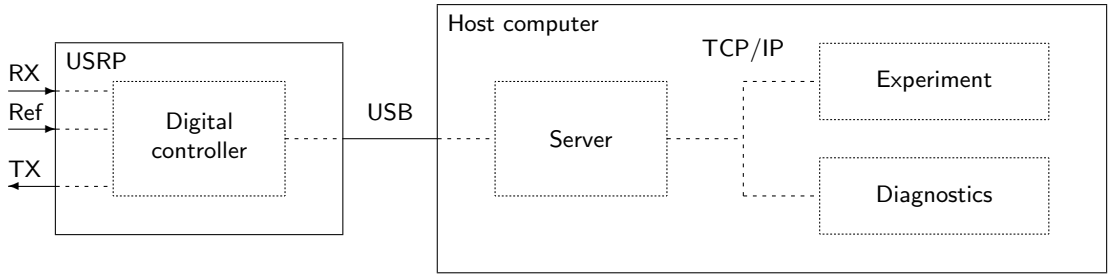
\begin{figure}
\begin{center} 
\setlength{\unitlength}{0.0100in}

\begin{picture}(600, 200)



\put(50,50){\framebox(150,100){}}
\put(55,145){\makebox(0,0)[tl]{\sf \footnotesize USRP}}

\put(80,70){\dashbox(90,60){}}
\put(100,90){\shortstack{\sf \footnotesize Digital \\
                        \sf \footnotesize controller}}


\put(250,30){\framebox(350,140){}}
\put(255,165){\makebox(0,0)[tl]{\sf \footnotesize Host computer}}

\put(280,70){\dashbox(90,60){\sf \footnotesize Server}}
\put(450,110){\dashbox(120,40){\sf \footnotesize Experiment}}
\put(450,50){\dashbox(120,40){\sf \footnotesize Diagnostics}}



\put(25,125){\vector(1,0){25}}
\put(25,130){\makebox(0,0)[lb]{\sf \footnotesize RX}}
\multiput(50,125)(5,0){6}{\line(1,0){2}} 

\put(25,105){\vector(1,0){25}}
\put(25,110){\makebox(0,0)[lb]{\sf \footnotesize Ref}}
\multiput(50,105)(5,0){6}{\line(1,0){2}}

\put(50,75){\vector(-1,0){25}}
\put(25,80){\makebox(0,0)[lb]{\sf \footnotesize TX}}
\multiput(50,75)(5,0){6}{\line(1,0){2}} 

\put(200,100){\line(1,0){50}}
\put(225,110){\makebox(0,0)[b]{\sf \footnotesize USB}}
\multiput(170,100)(5,0){6}{\line(1,0){2}} 
\multiput(250,100)(5,0){6}{\line(1,0){2}} 

\put(410,140){\makebox(0,0)[b]{\sf \footnotesize TCP/IP}}

\multiput(370,100)(5,0){8}{\line(1,0){2}} 
\multiput(410,70)(5,0){8}{\line(1,0){2}} 
\multiput(410,130)(5,0){8}{\line(1,0){2}} 
\multiput(410,70)(0,5){12}{\line(0,1){2}} 

\end{picture}
\caption{System hardware and software\label{fig:system-block}}
\end{center}
\end{figure}

Figure~\ref{fig:system-block} shows the cantilever controller system,
including the two main hardware components (solid boxes): the host
computer and the USRP (Universal Software Radio Peripheral), which
acts as the controller.  The USRP and host communicate over a USB 2.0
(Universal Serial Bus).  The USRP is in its own small case (21 cm
wide, 5 cm high, 14 cm deep) and the bus runs through a cable (up to 5
m, longer with extensions), so it is easy to connect the USRP to a
different host.

\subsection{USRP (Controller)}

All of the controller hardware including the FPGA, data converters,
timing, signal conditioning, and connectors are included in the
USRP~\cite{Ettus:USRP,GNURadio:USRP}.  The USRP is designed to support
software-defined radio, in particular the open-source GNU Radio
project~\cite{GNURadio:Trac}.  Its design is open source; its
schematics are included in the GNU Radio software distribution.  The
USRP is commercially available.

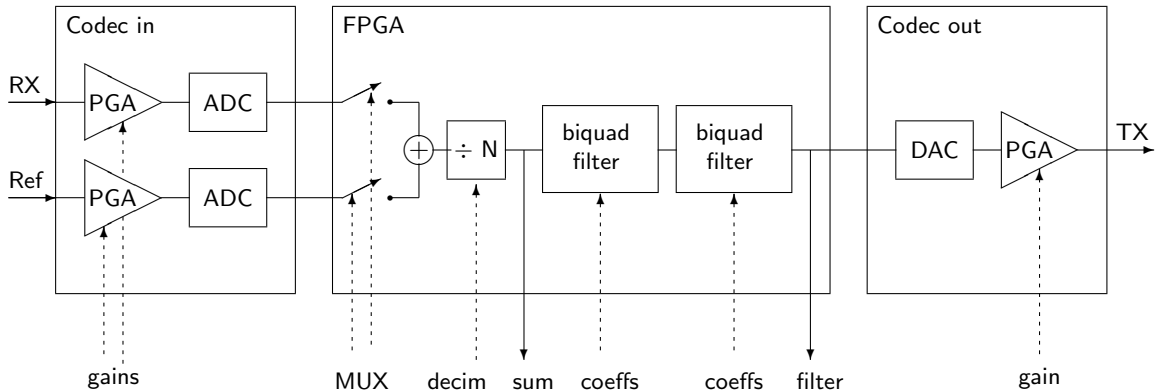
\begin{figure}
\begin{center} 
\setlength{\unitlength}{0.0100in}

\begin{picture}(600, 200)(0,-50)


\put(25,0){\framebox(125,150){}}
\put(30,145){\makebox(0,0)[tl]{\sf \footnotesize Codec in}}

\put(40,80){\line(0,1){40}}
\put(80,100){\line(-2,1){40}}
\put(80,100){\line(-2,-1){40}}
\put(42,100){\makebox(0,0)[l]{\sf \footnotesize PGA}}

\put(40,30){\line(0,1){40}}
\put(80,50){\line(-2,1){40}}
\put(80,50){\line(-2,-1){40}}
\put(42,50){\makebox(0,0)[l]{\sf \footnotesize PGA}}

\put(95,35){\framebox(40,30){\sf \footnotesize ADC}}
\put(95,85){\framebox(40,30){\sf \footnotesize ADC}}

\put(170,0){\framebox(260,150){}} 
\put(175,145){\makebox(0,0)[tl]{\sf \footnotesize FPGA}} 


\put(215,75){\circle{15}} 
\put(215,75){\makebox(0,0){$+$}}

\put(230,60){\framebox(30,30){\sf \footnotesize $\div$ N}} 

\put(280,53){\framebox(60,45){}} 
\put(290,65){\shortstack{\sf \footnotesize biquad \\          
                         \sf \footnotesize filter}}

\put(350,53){\framebox(60,45){}} 
\put(360,65){\shortstack{\sf \footnotesize biquad \\          
                         \sf \footnotesize filter}}

\put(450,0){\framebox(125,150){}}
\put(455,145){\makebox(0,0)[tl]{\sf \footnotesize Codec out}}

\put(465,60){\framebox(40,30){\sf \footnotesize DAC}}

\put(520,55){\line(0,1){40}}
\put(560,75){\line(-2,1){40}}
\put(560,75){\line(-2,-1){40}}
\put(522,75){\makebox(0,0)[l]{\sf \footnotesize PGA}}


\put(0,100){\vector(1,0){25}}  
\put(0,100){\line(1,0){40}}
\put(0,105){\makebox(0,0)[lb]{\sf \footnotesize RX}}

\put(0,50){\vector(1,0){25}}  
\put(0,50){\line(1,0){40}}
\put(0,55){\makebox(0,0)[lb]{\sf \footnotesize Ref}}

\put(560,75){\vector(1,0){40}}
\put(580,80){\makebox(0,0)[lb]{\sf \footnotesize TX}}

\put(50,30){\vector(0,1){5}} 
\multiput(50,35)(0,-5){14}{\line(0,-1){2}} 
\multiput(60,40)(0,-5){16}{\line(0,-1){2}} 
\multiput(60,90)(0,-5){6}{\line(0,-1){2}}  
\put(60,85){\vector(0,1){5}} 
\put(55,-50){\makebox(0,0)[b]{\sf \footnotesize gains}}

\put(80,100){\line(1,0){15}}
\put(80,50){\line(1,0){15}}

\put(135,100){\line(1,0){40}} 
\put(135,50){\line(1,0){40}}

\put(175,100){\vector(2,1){20}} 
\put(175,50){\vector(2,1){20}}

\multiput(180,52)(0,-5){17}{\line(0,-1){2}} 
\put(180,47){\vector(0,1){5}} 
\multiput(190,105)(0,-5){27}{\line(0,-1){2}} 
\put(190,100){\vector(0,1){5}} 
\put(185,-50){\makebox(0,0)[b]{\sf \footnotesize MUX}} 

\put(200,100){\circle*{3}} 
\put(200,50){\circle*{3}}

\put(200,100){\line(1,0){15}} 
\put(200,50){\line(1,0){15}}

\put(215,100){\line(0,-1){18}} 
\put(215,50){\line(0,1){18}}

\put(222,75){\line(1,0){10}} 

\put(260,75){\line(1,0){20}}

\multiput(245,57)(0,-5){19}{\line(0,-1){2}}  
\put(245,52){\vector(0,1){5}} 
\put(235,-50){\makebox(0,0)[b]{\sf \footnotesize decim}} 

\put(270,75){\vector(0,-1){110}} 
\put(275,-50){\makebox(0,0)[b]{\sf \footnotesize sum}} 

\multiput(310,52)(0,-5){17}{\line(0,-1){2}} 
\put(315,-50){\makebox(0,0)[b]{\sf \footnotesize coeffs}} 
\put(310,47){\vector(0,1){5}} 

\put(340,75){\line(1,0){10}}  

\multiput(380,52)(0,-5){17}{\line(0,-1){2}}  
\put(380,-50){\makebox(0,0)[b]{\sf \footnotesize coeffs}} 
\put(380,47){\vector(0,1){5}} 

\put(410,75){\line(1,0){55}} 

\put(420,75){\vector(0,-1){110}} 
\put(425,-50){\makebox(0,0)[b]{\sf \footnotesize filter}} 

\put(505,75){\line(1,0){15}}

\multiput(540,65)(0,-5){20}{\line(0,-1){2}} 
\put(540,-50){\makebox(0,0)[b]{\sf \footnotesize gain}}
\put(540,60){\vector(0,1){5}} 

\end{picture}
\caption{Controller signal path\label{fig:controller-block}}
\end{center}
\end{figure}

Figure~\ref{fig:controller-block} shows the signal path through the
USRP: the signal travels from left to right (horizontal solid lines),
digitized control parameters from the host computer enter on the USB
bus (vertical dotted lines) and the digitized acquired signals for the
host computer exit on the USB (vertical solid lines).  The
multiplexer, decimator, and filters execute on the FPGA, an Altera
Cyclone EP1C12, which provides 12,060 logic elements (each comprising
several logic gates) and 239,616 bits of
storage~\cite{Altera:Cyclone}.  The data converters on the USRP are
included in two Analog Devices AD9862 codecs, each with two analog to
digital converters (ADC) and two digital to analog converters
(DAC)~\cite{AD:9862}.  The ADCs have 12 bit resolution; the
DACs have 14 bit resolution.  The AD9862 also includes
programmable gain amplifers (PGA) at the input of each ADC
and the output of each DAC.  The two blocks labelled ``Codec
in'' and ``Codec out'' in Fig.~\ref{fig:controller-block} represent
different parts of a single AD9862 device (there is one additional DAC
with PGA which we do not use).

The ADC output pins are connected directly to FPGA input pins, so the
digitized signal reaches the FPGA with less latency than would be
possible if the ADCs were connected through a bus shared by several
devices (as is common in microprocessors, including DSPs).

The ADCs sample at 64 MHz and the FPGA is clocked at 64 MHz.
The DAC samples at 128 MHz.  The FPGA program includes a
decimator ($\div N$ in Fig.~\ref{fig:controller-block}) which can
divide down the sampling frequency used by the filters.  The precision
required for the filter coefficients increases as the sampling
frequency grows relative to the resonant frequency of the filter.  A
programmable sampling frequency makes it possible implement filters
that work well over a wide range of signal frequencies, even when the
filter coefficients are represented with limited precision.  A good
compromise for a filter with resonance near 8 kHz divides the 64 MHz
clock by 128 to sample at 500 kHz.

The FPGA, codecs, and USB hardware are on the USRP motherboard.  The
USRP also includes two daughterboards, LFRX and LFTX (Low-Frequency
Receiver and Transmitter, where low frequency means below 30 MHz).
These provide connectors and AD813x differential amplifiers at the
inputs and outputs for buffering and signal
conditioning~\cite{AD:813x}.  The amplifier circuits are fixed at
unity gain and 50 ohm impedance.  The daughterboards have no
programmable elements (daughterboard components do not appear in
Fig.~\ref{fig:controller-block}).

The signal path shown in Fig.~\ref{fig:controller-block} only uses half
the USRP: one of two AD9862 codecs, and two of four daughterboards.
Another independent signal path could be accommodated in the same
USRP.   Both signal paths would execute in parallel so both would run
at full speed.  

\subsection{Host computer}

The host computer stores the controller program that runs on the FPGA.
Before an experiment, a program on the host computer loads the
controller program into the FPGA.  During an experiment, the host
computer executes experiment control software that can set and
adjust controller parameters and acquire the measured signals.

\section{Software}

\subsection{Controller software}

The controller program comprising the multiplexer and digital filter
executes on the FPGA in the USRP.  

The multiplexer selects among input signals to connect to the digital
filter.  It can connect the cantilever position signal or a reference
signal (RX or Ref in Fig.~\ref{fig:controller-block}), or neither, or
their sum.  All four combinations are used in our calibration protocols.

The filter consists of two second-order stages, also called {\em
biquadratic} stages, commonly called ``biquads''.  Each stage
implements a transfer function represented by a ratio of second-order
polynomials, with three coefficients in both the numerator and
denominator.  The signals and coefficients are 24-bit integers.
Using biquad stages requires less precision than a single
higher-order filter.

Two biquad filter stages are sufficient to implement transfer
functions for optimal cantilever controllers derived by the method of
the appendix.  The transfer functions are computed
and then translated to biquad stages by software on the host computer
(section~\ref{sect:host-sw}).

In addition to the controller itself, the FPGA also executes code that
communicates with the host computer over the USB.  The bus control
code, and several modules used by the controller, are from the 
open-source GNU Radio project~\cite{GNURadio:Trac}.

The controller program is written in the Verilog hardware description
language (HDL)~\cite{Bhasker:98}.  The several Verilog source files
are compiled or {\em synthesized} into a single binary file called a
{\em bitstream} which is stored on the host computer and loaded into
the FPGA on command.  An FPGA can be reprogrammed by loading
another bitstream.  The host computer can store several bitstreams
that provide different controller capabilities and load them into the
FPGA under program control.


The {\em word length} is the number of bits used to represent numbers
in a program.  An FPGA program can specify any word length at each
stage in each computation, unlike a conventional processor or DSP
where only a small fixed set of word lengths is supported by the
machine arithmetic.  FPGA capacity is limited, so an important
design task is to choose the minimum word lengths that provide
sufficient accuracy.  Different word lengths can be chosen for the
signal path (solid lines in Fig.~\ref{fig:controller-block}) and for
the filter coefficients (dotted lines near the center of
Fig.~\ref{fig:controller-block}).

If the word length of the signal path is too small, the filter
exhibits a nonlinear input/output relation with a threshold at small
amplitudes and saturation at large amplitudes.  With a 16-bit signal
path, the linear range is only from about 0.1 V to 0.5 V input
amplitude, which is too little dynamic range for our experiments.  The
word length required for the filter coefficients increases as the
sampling frequency grows relative to the resonant frequency of the
filter.  If the word length of the coefficients is too small, the
actual resonant frequency of the filter differs from the intended
(computed) value. A good compromise for a filter with resonance near 8
kHz divides the 64 MHz clock by 128 to sample at 500 kHz.  Under
these conditions, 16 bit coefficients can result in an actual resonant
frequency that differs about 16 Hz from the intended value, which is
too much for our experiments.  We found that 24 bits were adequate for
both signal path and coefficients; some intermediate results use word
lengths up to 50 bits.  We found the acceptable word lengths by trial
and error, although did do some simulations in
MATLAB~\cite{Mathworks:FixedPoint} to confirm that the symptoms we
observed were explained by the limited word length, rather than
programming errors or some other cause.


We have only very limited ability to observe and debug synthesized
Verilog code executing on the FPGA.  Therefore we do most debugging in
simulation.  First we check the design in MATLAB, as described above.
Then on the host computer we write the Verilog code and use the Icarus
Verilog compiler to build an executable simulation~\cite{Icarus}.  We
execute the simulation on the host and use the GTKWave viewer on the
host to show its results in a display that resembles an oscilloscope
or logic analyzer~\cite{GTKWave}.  Finally we use a synthesizer to
compile the debugged Verilog to a bitstream that can be executed on
the FPGA~\cite{Altera:Quartus}.

The bitstream for this controller program uses 2,765 logic elements
and 66,304 memory bits in the FPGA, about one fourth of the
available resources.  Another independent signal path could be accommodated in
the same FPGA, or higher-order transfer functions could be implemented
by adding more biquad filter stages.

The controller software is open source and is available from the
project web site~\cite{Jacky:08}.

The measured performance of the controller is discussed in
section~\ref{sect:performance}.

\subsection{Host computer software} \label{sect:host-sw}

The experiment software including calibration protocols, data
acquisition and storage, calculation of the controller transfer
function, etc., executes on the host computer.

The host software is organized as a client and server.  The client
comprises most of the experiment control, while the server is a small
component that copes with the details of the controller and its
interface.  For example, the client calculates the transfer function
for the optimal cantilever controller as described in
the appendix.  The transfer function is represented
as a ratio of polynomials with four numerator and four denominator
coefficients.  The server transforms this transfer function for two
cascaded biquad sections by factoring the polynomials, sorting and
collecting factors, then multiplying them in two groups.  It also
replaces floating point coefficients with scaled 24-bit integers.  It
loads the integer coefficients into particular registers in the FPGA
using the USB.  The client software does not depend on
controller details such as filter architecture, hardware organization,
number representation, and connection technology; these only affect
the server.  Therefore the same server could be used with different
experiments; the particulars of each experiment are encoded in the
client.  Thanks to this organization, our client software
is largely unchanged from earlier versions that used completely
different controller technology based on a DSP.

Client and server communicate by sending text messages over a TCP/IP
socket connection.  In Fig.~\ref{fig:controller-block}, both execute
on the same computer.  They can also execute on different computers
(we have used both configurations).

\begin{figure}
\begin{center} 
\includegraphics[width=4in]{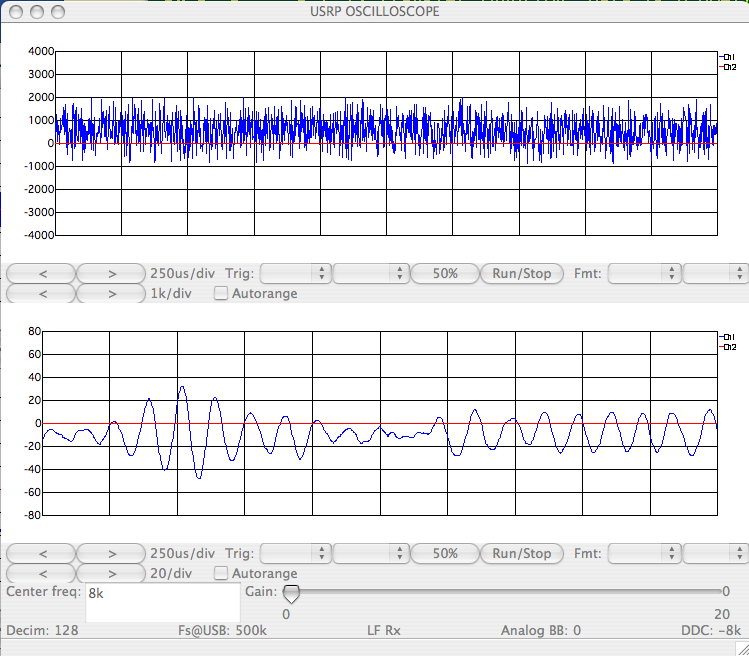}
\caption{Server program oscilloscope-like display\label{fig:usrp-oscope}}
\end{center}
\end{figure}

The server also performs data acquisition, acquiring the stream of
digitized samples from the FPGA at two points in the signal path: the
outputs of the multiplexer and the digital filter (labeled ``sum'' and
``filter'' in Fig.~\ref{fig:controller-block}).  The server can
optionally display these data as a spectrum or time series on an
oscilloscope-like display, as shown in Fig.~\ref{fig:usrp-oscope}.

The server is written in Python and uses libraries from the open
source GNU Radio and SciPy projects~\cite{Oliphant:07,GNURadio:Trac,SciPy}.

Our usual client is a large experiment control program written in
LabView~\cite{NI:LabView}.  We also wrote a small client in Python for
diagnostics and testing.  Both clients can run at the same time,
because both open and close the socket connection each time they send
a message.

The host software is open source and is available from the project web
site~\cite{Jacky:08}.

\section{Performance} \label{sect:performance}

\begin{figure}
\begin{center} 
\includegraphics[width=5in]{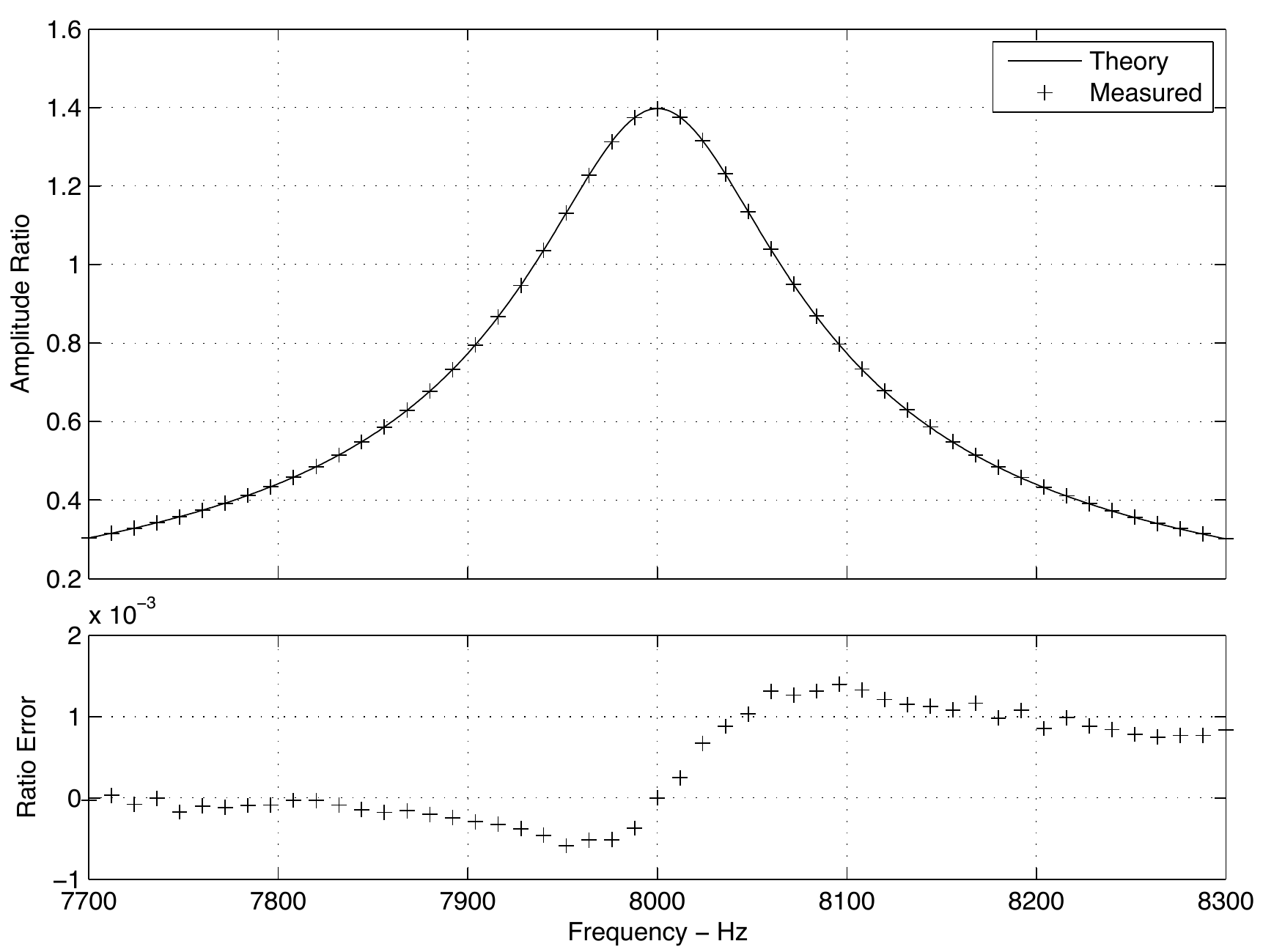}
\caption{Calculated and measured transfer functions: amplitude
\label{fig:transfer-function-ampl}}
\end{center}
\end{figure}
\begin{figure}
\begin{center} 
\includegraphics[width=5in]{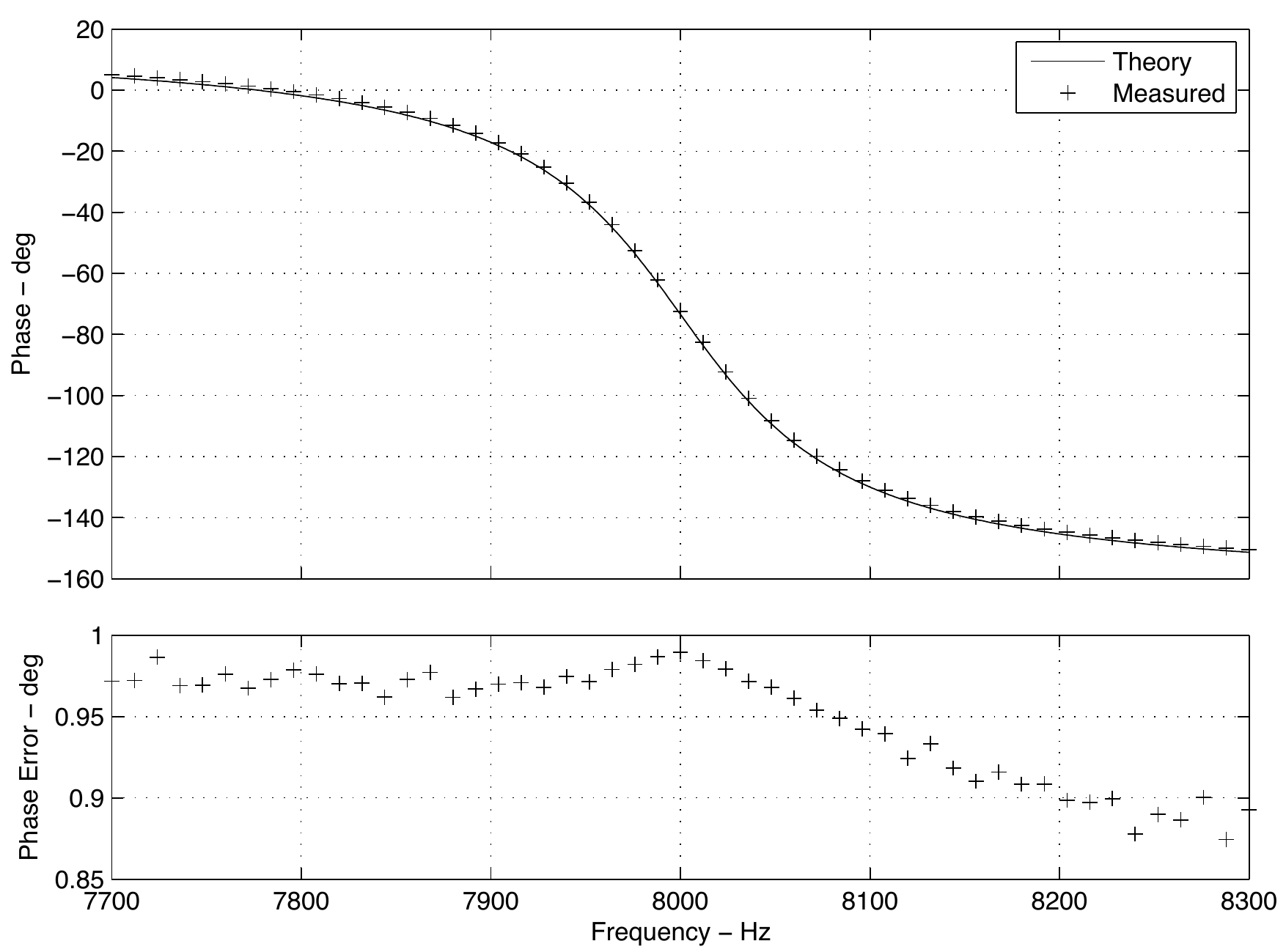}
\caption{Calculated and measured transfer functions: phase
\label{fig:transfer-function-phase}}
\end{center}
\end{figure}

We compared a computed transfer function to the transfer function
infered from measurements of the analog signals at the input and
output of the controller, and found close agreement
(Figs.~\ref{fig:transfer-function-ampl}
and~\ref{fig:transfer-function-phase}).  For this test we used a
controller transfer function that we had computed for a typical MRFM
experiment, by the method of the appendix, where the cantilever had
resonant frequency near 8000 Hz and an uncontrolled resonant quality
$Q$ near 10,000.  The transfer function is described by the four $b$
(numerator) and four $a$ (denominator) coefficients in
table~\ref{table:transfer-function}.

\begin{table}[h]
\begin{tabular}{|l|r|r|r|r|}\hline
$b_i$ &
$7.026189 \cdot 10^{-5}$ &
$1.027999 \cdot 10^{-4} $ &
$-5.927540 \cdot 10^{-5}$ &
$-9.181339 \cdot 10^{-5}$  \\ \hline
$a_i$ &
$1.000000 \cdot 10^0$ &
$-2.848528 \cdot 10^0$ &
$2.708790 \cdot 10^0$ &
$-8.588522 \cdot 10^{-1}$ \\ \hline
\end{tabular}
\caption{Controller transfer function (index $i$ ranges from 0 to 3,
  left to right)\label{table:transfer-function}}
\end{table}

Plots of the computed transfer function appear as solid lines in
Figs.~\ref{fig:transfer-function-ampl} (amplitude)
and~\ref{fig:transfer-function-phase} (phase).  At the resonant
frequency, the ratio of the output amplitude to the input amplitude is
about 1.4, and the output phase lags the input phase by about 80
degrees.  The server software factors and scales this transfer
function into two cascaded biquad sections with 24-bit integer
coefficients (table~\ref{table:filter-coeffs}).

\begin{table}[h]
\begin{tabular}{|l|r|r|r||r|r|r|}\hline
 & \multicolumn{3}{|c|}{Section 0}
 & \multicolumn{3}{|c|}{Section 1} \\ \hline \hline
$b_i$ & 35158 & 2293 & -32865 & 35158 & -49146 & \hspace*{0.3in} 0 \\ \hline
$a_i$ & -4194304 & 8339278 & -4187298 & -4194304 & 3608314 &  \hspace*{0.3in} 0 \\ \hline
\end{tabular}
\caption{Controller filter coefficients 
(in each section index $i$ ranges from 0 to 2, left to right)
\label{table:filter-coeffs}}
\end{table}

We measured the controller transfer function with a lock-in
amplifer~\cite{SRS:850}.  The SINE OUT signal from the lock-in was
applied to the input to the controller (RX in
figs.~\ref{fig:system-block} and~\ref{fig:controller-block}).  The
experiment software commanded the lock-in amplifier to generate sine
waves from 7700 to 8300 Hz at 100 Hz intervals.  The sine wave
amplitude was adjusted to 0.100 V.  The output from the controller (TX
in the figures) was applied to the SIGNAL IN A-I input of the lock-in.
At each frequency, the in-phase and quadrature components X and Y of
the controller output were calculated by the lock-in amplifier, with a
time constant of 3 sec, and the experiment software calculated the
amplitude and phase of the transfer function from 
X and Y samples.  The measured transfer function points appear as
crosses in Figs.~\ref{fig:transfer-function-ampl} (amplitude)
and~\ref{fig:transfer-function-phase} (phase).  The errors
(differences between calculated and measured values) appear in the
plots below the transfer functions.  The amplitude ratio error is less
than 0.002, and the phase error is less than one degree, over the
entire measured frequency range.  These errors are acceptable in our
experiment.

The computed and measured transfer functions have a phase lag of about 80
degrees at the resonant frequency.  The computed transfer function
includes phase compensation to account for the phase lag
contributed by various components in the entire signal path
(appendix, Fig.~\ref{Implementation Fig} and 
section~\ref{sect:phase}).  The lag contributed by the fixed
latency of the digital controller was estimated by applying a 50 Hz
square wave to the input and observing (on an oscilloscope) the delay
until the output starts ringing.  This showed a latency of about 2
usec, similar to the sampling interval (at the 500 kHz sampling
frequency), so the digital controller contributes about 6 degrees of
lag at 8000 Hz.

The gains of the PGAs (Fig.~\ref{fig:controller-block}) were set to to
1, nominally.  This results in (nearly) equal amplitudes for RX and
TX, when a controller with unity gain is loaded.  The PGAs are not
precisely calibrated so the overall gain is not exactly 1.0.  It is
possible to trim the output PGA (via software) to obtain an overall
gain very near 1.0, but we choose instead to simply keep track of the
observed overall gain in our experiment software.  The observed
overall gain in these measurements was 1.038.

\section{Application}

\begin{figure}
\begin{center} 
\includegraphics[width=6in]{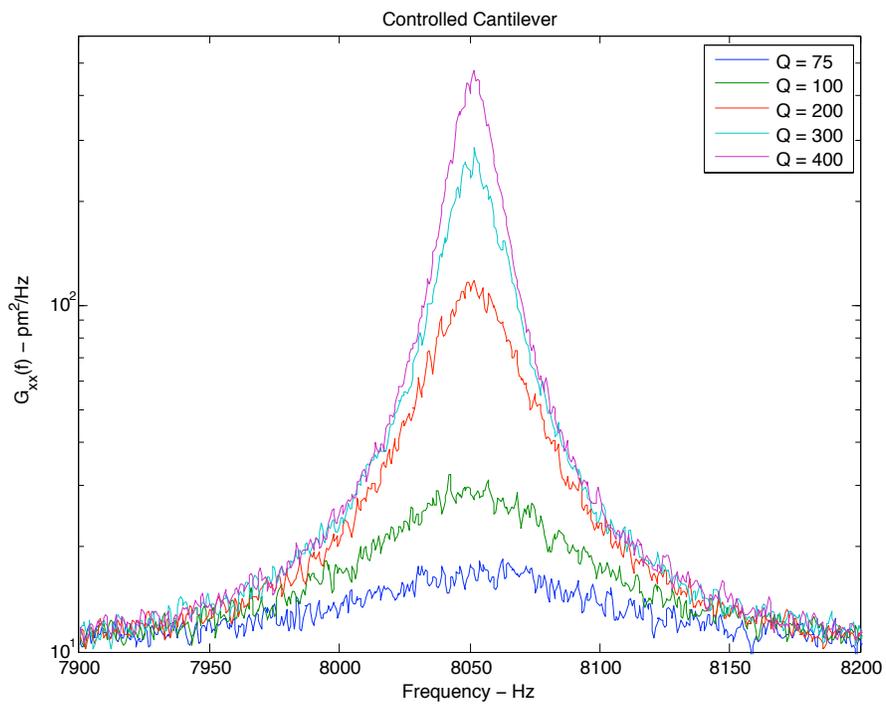}
\caption{Controlled cantilever power spectra\label{fig:cantilever-spectra}}
\end{center}
\end{figure}

We used the controller in an experiment with a cantilever with a
resonant frquency of 8347 Hz and an uncontrolled $Q$ of about 10,000.
Figure~\ref{fig:cantilever-spectra} shows the one-sided power spectrum
of the noise measured (by a spectrum analyzer~\cite{SRS:780}) from the
controlled cantilever with several different controller transfer
functions, with controlled $Q$ values 75, 100, 200, 300 and 400.
These measurements show the expected results.

\section{Discussion}

We anticipate adapting the platform described here to different
control schemes.  The following paragraphs describe one that we are
considering.

A goal of MRFM is to achieve atomic-scale resolution by resolving
individual nuclear spins.  A promising approach to this goal would use
cantilevers that oscillate at radio frequencies (1 -- 10
MHz)~\cite{Sidles:92b}.  Such high cantilever frequencies press the
performance limits of available digital control technologies.  An
alternative uses \emph{heterodyne control}, where the central
controller elements process signals at much lower frequencies than the
cantilever oscillation~\cite{Kriewall:06}.  It confers another
potential advantage because a heterodyne controller might also act as
the lock-in detector for the MRFM signal, eliminating the separate
lock-in amplifier we now use in our experiments~\cite{Chao:04}.

In earlier work, we demonstrated a digital heterodyne controller with
an emulated radio frequency cantilever~\cite{Kriewall:06} (our
emulation was a real-time digital simulation with analog inputs and
outputs that resembled the signals expected from a radio frequency
cantilever). We have begun experimenting with an FPGA implementation
of a heterodyne controller~\cite{GNURadio:MRFM}.  The hardware and
host computer software is the same as described here.  The only
difference is the FPGA code for the controller, which implements a
different block diagram than the one shown here 
in Fig.~\ref{fig:controller-block}.

\begin{acknowledgments}
This work was supported by the Army Research Office
(ARO) Multi-University Research Initiative (MURI) W911NF-05-1-0403.
\end{acknowledgments}

\appendix*  

\section{Controller transfer function} \label{app:transfer-fcn}

Here we show how to calculate filter coefficients for an optimal
controller from measured cantilever characteristics.  We derive the
controller transfer function, following the theory of optimal feedback
control of force cantilevers in~\cite{Garbini:96}.  After reviewing
briefly the continuous-time specification of that controller, we
derive a discrete-time equivalent.
Practical hardware and software issues, along with 
compensation of parasitic effects, are treated.

\subsection{Continuous-time Control} \label{sec:continuous} 

\begin{figure}
	\centering 
	\includegraphics[width=4in]{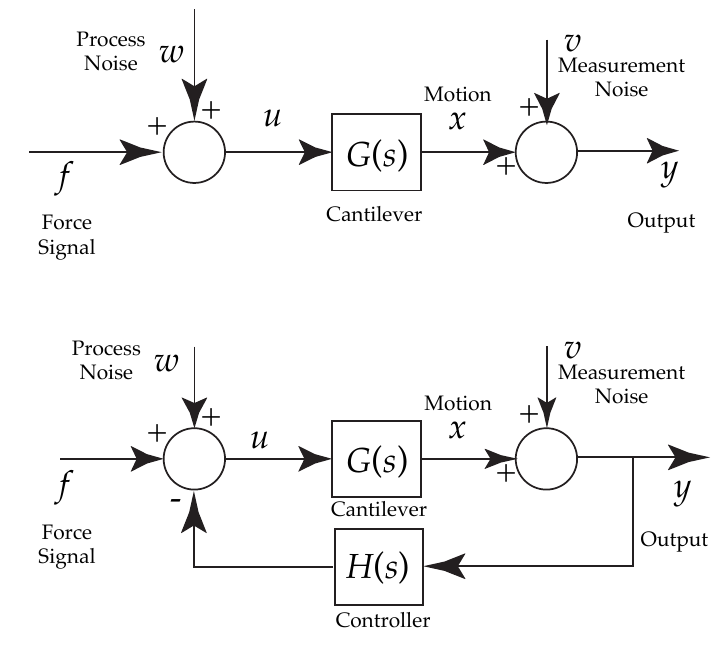} 
	\vspace{-.1in}
	\caption{Open and closed loop dynamic systems} 
	\label{SNR Picture} 
	\end{figure}

In Fig.\ \ref{SNR Picture}, a linear dynamic feedback control element 
is added to the open loop system.  The dynamic relationship between 
the forces $u$ applied to the cantilever tip and the motion $x$ of the 
tip is represented by the transfer function $G(s)$.  In addition, $f$ 
is the force signal to be detected, $w$ is additive process noise 
force, $v$ is measurement noise added to the actual position of the 
cantilever, and $H(s)$ is the transfer function of the feedback 
controller.

Many techniques are available for the design of $H(s)$.  All require 
knowledge of the physical characteristics of the cantilever, the noise 
processes $w$ and $v$, and a design goal for controller performance.  
Below, the techniques of optimal control and estimation are combined 
to determine the continuous-time transfer function of the controller 
$H_{oc}¥(s)$ which relates measured motion of the cantilever $y$ to 
the control component of the input force $u$ (Eq.\ (31) of \cite{Garbini:96}).

	\begin{equation} 
	H_{oc}(s)={U(s) \over Y(s)}={{K_{oc}(s+z_{oc})} \over {s^2+{{\omega_{oc}} \over
	Q_{oc}}s+\omega_{oc}^2}} \ ,
	\label {CLaw}
	\end{equation}
where
	\begin{displaymath} 
	\nonumber{{K_{oc}} \over \omega_n}={1 \over 2}k\alpha\beta(\alpha+\beta)+ {{k\alpha\beta}
	\over {Q}} \ ,
	\end{displaymath}
	
	\begin{displaymath} 
	{z_{oc} \over \omega_n}={{{{\alpha\beta} \over 2}-2+{{\alpha+\beta} \over Q}+ {2 \over
	{Q^2}}} \over {\alpha+\beta+{2 \over Q}}} \ ,
	\end{displaymath}

	\begin{displaymath} 
	{{\omega_{oc}^2} \over \omega_n^2}={{(\alpha+\beta)}^2 \over 2} +
	{{\alpha+\beta} \over {Q}}+1 \ ,
	\end{displaymath}
	
	\begin{displaymath} 
	{Q_{oc}}={{\sqrt{{{(\alpha+\beta)^2} \over 2} + {{\alpha+\beta} \over
	{Q}}+1}} \over {\alpha+\beta+{1 \over Q}}} \ .
	\end{displaymath}
where $k$, $\omega_{n}$, and $Q$ are the stiffness, natural frequency, 
and resonant quality of a single-mode cantilever model.  The 
parameters $\alpha$ and $\beta$ are determined from the control and 
estimation requirements.

The value of $\alpha$ is defined by the need to keep the system 
stationary within acceptable deviations from $x=0$, while using acceptable 
amounts of control $u$.

\begin{equation} \alpha=\sqrt{{1 \over{Q^2}}+2\xi^2}-{1
	\over Q}\ \hbox{and}\ \xi^2=\sqrt{1+{U \over{k^2 X}}}-1 \ .
	\end{equation} 
$\alpha$ is a function of the maximum allowed variances of the displacement 
$X=E[x_m^2]_{max}$ and the control force $U=E[u^2]_{max}$.  The 
definitions of $\xi$ and $\alpha$ indicate that the relevant physical 
parameter is the ratio of the maximum control force to the force that 
would be caused in a spring subjected to the maximum displacement.

The estimator parameter $\beta$ represents the quality of the data 
used to estimate the system state, and is determined by the relative 
amounts of process and measurement noise.

	\begin{equation} 
	\beta=\sqrt{{1\over{Q^2}}+2\psi^2}-{1\over Q}\ \hbox{and}\ \psi^2=\sqrt{1+{W\over
	{k^2V}}}-1 \ .
	\end{equation} 
where constants $W$ and $V$ are the power spectral densities of 
stationary, zero-mean, Gauss-Markov random processes.  Large 
measurement noise means that the observations are weighted less in the 
estimate of the state, while small measurement noise allows large 
gains in the estimator.
	
The controller has two complex poles and a real zero.  As the 
allowable control effort is increased (greater $\alpha$), or the 
measurement noise is reduced (greater $\beta$), the natural frequency 
of the controller rises and its $Q_{oc}$ is diminished.  Again, in the 
limit as $(\alpha+\beta)$ approaches infinity, $Q_{oc}$ approaches 
${\sqrt{2}/2}$.

Ultimately, the control accuracy can be no better than the inferred 
state.  The control value, $\langle u^2\rangle$, approaches infinity 
as $\alpha\rightarrow\infty$.  However, even when large control effort 
is available, little is gained by allowing $ \alpha$ to be arbitrarily 
large.  The poles of the closed loop system are the union of the poles 
of the optimal controller and estimator.  Since poles of smaller 
magnitude will tend to dominate the dynamic response, a practical 
guide is to make $ \alpha$ smaller than $\beta$, but of similar 
magnitude.

While the closed-loop system is of fourth order, it is often useful to 
describe the controller design goal in terms of the effective resonant 
quality $Q_{cl}$ of the dominant closed-loop poles.  Assuming that 
$\alpha$ is smaller than $\beta$, the value of $\alpha$ corresponding 
to the closed-loop quality $Q_{cl}$ is given by (Eq.\ (16) of 
\cite{Garbini:96})

	\begin{equation} 
	\alpha={1\over Q}\left 
	[1-\sqrt{{2Q^{2}-1\over{2Q_{cl}^{2}-1}}}\ \right 
	]\simeq{{1}\over{Q_{cl}}}-{{1}\over{Q}} .
	\end{equation} 
Alternately the design goal can be stated in terms of the closely 
related quantities:  the bandwidth 
\mbox{$\Delta\omega_{cl}=\omega_{n}/Q_{cl}$}, or 
the damping time $\tau_{cl}=2Q_{cl}/\omega_{n}$.

Finally, the performance of the optimally controlled closed loop 
system in terms of the mean-square cantilever displacement and control 
force can be predicted from (Eqs.\ 38-39 of \cite{Garbini:96})

	\begin{equation} 
	{{\langle x_m^2\rangle} \over \omega_n}= {{V\beta\bigl[ {{2\alpha^2+\beta^2}
	\over 2}+{{2\alpha+\beta}\over Q}+\alpha\beta+2 \bigr]\bigl[ {{2\alpha+\beta}
	\over 2}+{1 \over Q} \bigr]} \over {\bigl[ \alpha+{1 \over Q} \bigr]\bigl[
	\alpha^2+{2\alpha \over Q}+2 \bigr]}} \ ,
	\label{Xperf}
	\end{equation} 
and (Eq.\ (39) of \cite{Garbini:96})

	\begin{eqnarray} 
	{{\langle u^2\rangle} \over \omega_n}&=& {V(k\alpha\beta)^2 \over {8(\alpha+{1
	\over Q})(\alpha^2+{{2\alpha} \over Q}+2)}} \times\nonumber\\
	& &\Bigl[{ \alpha^4+ 2\alpha^3\beta+ 2\alpha^2+ 2\beta^2
	+ {3 \over 2}\alpha^2\beta^2+ 8+} \nonumber\\
	& &{ {{6\alpha^3+10\beta\alpha^2+4\beta^2\alpha} \over Q}+
	{{2\beta^2-8+14\alpha^2+16\alpha\beta} \over {Q^2}}+} \nonumber\\
	& &{  {{8\alpha+16\beta} \over
	{Q^3}}+ {8 \over {Q^4}} }\Bigr] \ .
	\label{Uperf}
	\end{eqnarray} 
Since $\beta$ is a function of $V$, it is useful to express $V$ in 
Eqs.\ (\ref{Xperf}) and (\ref{Uperf}) in terms of the (experimentally measured) 
\emph {uncontrolled} (open loop) rms amplitude $\langle {y}_{ol}^{2} \rangle^{1 \over 2}$
	\begin{equation} 
	V={
	8\langle {y}_{ol}^{2} \rangle^{1 \over 2}Q/\omega_{n}
	\over 
	(\beta^{4}+4\beta^{2})Q^{2}+4(4{\Delta\omega_{V} \over 
	\pi\omega_{n}}+2\beta+\beta^{3})Q+4\beta^{2}}
	\end{equation} 
where $\Delta\omega_{V}$ is the effective measurement noise bandwidth.  

\subsection{Discrete-time Control} \label{sec:discrete} 

The digital controller shown in Fig.~\ref{fig:controller-block}
implements the feedback element $H(s)$.  
The digital controller periodically samples
the voltage signal corresponding to the measured cantilever motion,
computes the appropriate control law, and imposes the result through a
zero-order hold at its output.  In our case, the control law is a
discrete-time equivalent of the continuous-time transfer function in
Eq.\ (\ref{CLaw}).

In general, the control law implemented by the digital controller can
be any linear causal time-invariant difference equation of the form

	\begin{equation} 
	y(n)= - \sum_{k=1}^{N}\frac{a_{k}}{a_{0}}y(n-k) 
              + \sum_{k=0}^{M}\frac{b_{k}}{a_{0}}x(n-k)
	\label{DiffEq}
	\end{equation} 
where $y(n)$ and $x(n)$ are samples of the output and input 
respectively, and $a_{k}$ and $b_{k}$ are constants.

Before we derive the difference equation several hardware and software 
issues must be considered.

\subsection{Implementation Considerations} \label{sec:implementation} 

\begin{figure} 
	\centering 
	\includegraphics[width=5in]{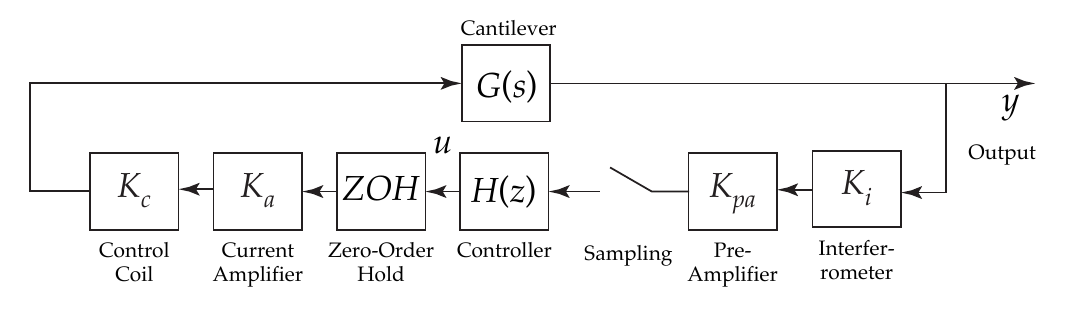} 
	\newline
	\caption{Implementation block diagram} 
	\label{Implementation Fig} 
	\end{figure}

Figure\ \ref{Implementation Fig} shows the complete cantilever control 
system, including the elements of cantilever motion sensing, force 
actuation, conversion and amplification.  
Below, we consider the specifications and characteristics of each 
block.  Throughout, we assume that the A/D and D/A converters are 
linear with inversely related gains: $K_{ad}=1/K_{da}$.

\subsubsection{Sampling Rate}
Faithful discrete-time reproduction of the continuous-time controller 
requires that the sampling rate be large in comparison with the 
natural frequency of the system.  A good rule-of-thumb is to select a 
sampling frequency at least 30 times larger than the cantilever 
natural frequency.  Typically, our sampling rate is 0.5 MHz for a 
natural frequency of 7500 Hz.

\subsubsection{The interferometer}
The fiber optic interferometer converts the displacement of the 
cantilever into a proportional voltage signal.  A typical sensitivity 
for our interferometer is \mbox{$K_{i}=10\times 10^{6}$ V/m}.  The 
interferometer bandwidth is limited to approximately 100 kHz.

\subsubsection{Force Actuator}
Magnetic actuation provides the restoring feedback force to the 
cantilever.  The force is proportional to the current passing through 
the control coil.  The actuator can be calibrated by driving the coil 
(open loop) with a sinusoidal current of known amplitude at the 
cantilever natural frequency, and measuring the resulting cantilever 
motion.  From that measurement and the known stiffness, quality, and 
natural frequency of the cantilever, the actuator sensitivity $K_{c}$ 
is determined.  

A typical actuator gain is \mbox{$K_{c} = 35\times 
10^{-12}$ N/A}.

\subsubsection{Amplification}
Two stages of amplification are used to match the dynamic ranges of 
the interferometer and coil actuator to the input/output ranges of 
the digitial controller.

Suppose that the full-scale ranges of both the A/D and D/A converters 
are $V_{fs}$.  A simple strategy is to select the interferometer 
pre-amplifier gain $K_{pa}$ such that the predicted rms value of the 
controller input is one third of the A/D converter range:

	\begin{equation} 
	K_{pa}={V_{fs} \over 3K_{i}\sqrt{{\langle x_m^2\rangle}+{\Delta\omega_{n}V 
	\over \pi }}}
	\end{equation} 

Then select the current amplifier gain $K_{a}$ such that the predicted rms 
value of the controller output is one third of the D/A converter range:

	\begin{equation} 
	K_{a}={ 3{\langle u^2\rangle}^{1 \over 2} \over K_{c}V_{fs}}
	\end{equation} 

These rules ensure that signals substantially fill the A/D and D/A 
converter ranges (thereby minimizing quantization noise), while only 
rarely allowing the signals to exceed those ranges.

For our system, the converter ranges are \mbox{$V_{fs}=$ 1 V}.  Typical gain 
values are \mbox{$K_{pa}= 14.4$ V/V}, and \mbox{$K_{a}= 0.001$ A/V}.

\subsubsection{Phase-lead Compensation} \label{sect:phase}
The optimal controller described in Eq.\ (\ref{CLaw}) was derived under 
the assumption that the remainder of the control loop in Fig.\ 
\ref{SNR Picture} consists only of the single-mode (2nd order) 
cantilever model.  As we have seen, any practical implementation 
includes additional dynamic components.  The primary effect of these 
components is to introduce additional phase lag within the relatively 
narrow bandwidth of the controlled cantilever.  Generally, phase lag 
tends to lower the damping, causing the performance to deviate from 
the predictions of Eqs.\ (\ref{Xperf}) and (\ref{Uperf}).

\pagebreak Sources of additional phase lag include:
\begin{itemize}
\item Finite bandwidths of the interferometer, its pre-amplifier, and 
the actuator coil current amplifier.  Each device is characterized by 
low gain and large phase lag at high frequencies.  Even components 
with bandwidths exceeding 0.1 MHz, will each contribute a few degrees 
of phase lag at the cantilever frequency.
 
\item The A/D conversion delay $T_{d}$ causes an effective phase lag of 
$\omega T_{d}$ radians.

\item Similarly, the D/A converter zero-order hold, of sample time 
$T$, adds $\omega T/2$ radians of phase lag.
\end{itemize}

While analytical predictions are useful, the most reliable method of 
determining the \emph{total} phase lag is by careful measurement of 
the actual system.  Minor phase lag will not seriously alter the 
system behavior and, in many cases, can be ignored.  Of course, we 
could have accounted for the transfer function of each of these 
additional components during the design of the controller.  However, a 
closed-form solution for the controller would not be possible, and the 
insight afforded by Eq.\ (\ref{CLaw}) would be lost.  Instead, we can 
easily compensate for the phase lag by augmenting the control law with 
a phase-\emph{lead} dynamic component.

Consider the following simple phase-lead transfer function

	\begin{equation} 
	G_{c}(s)={(1+\eta\tau s) \over \sqrt{\eta}(1+\tau s)}
	\end{equation} 
where

	\begin{eqnarray} 
	\eta={1+\sin{\phi_{m}} \over 1-\sin{\phi_{m}}}&\textup{\hspace{.2in}and\hspace{.2in}}& 
	\tau={1 \over \omega_{m}\sqrt{\eta}}\nonumber
	\end{eqnarray} 

\begin{figure} 
	\centering 
	\includegraphics[width=4in]{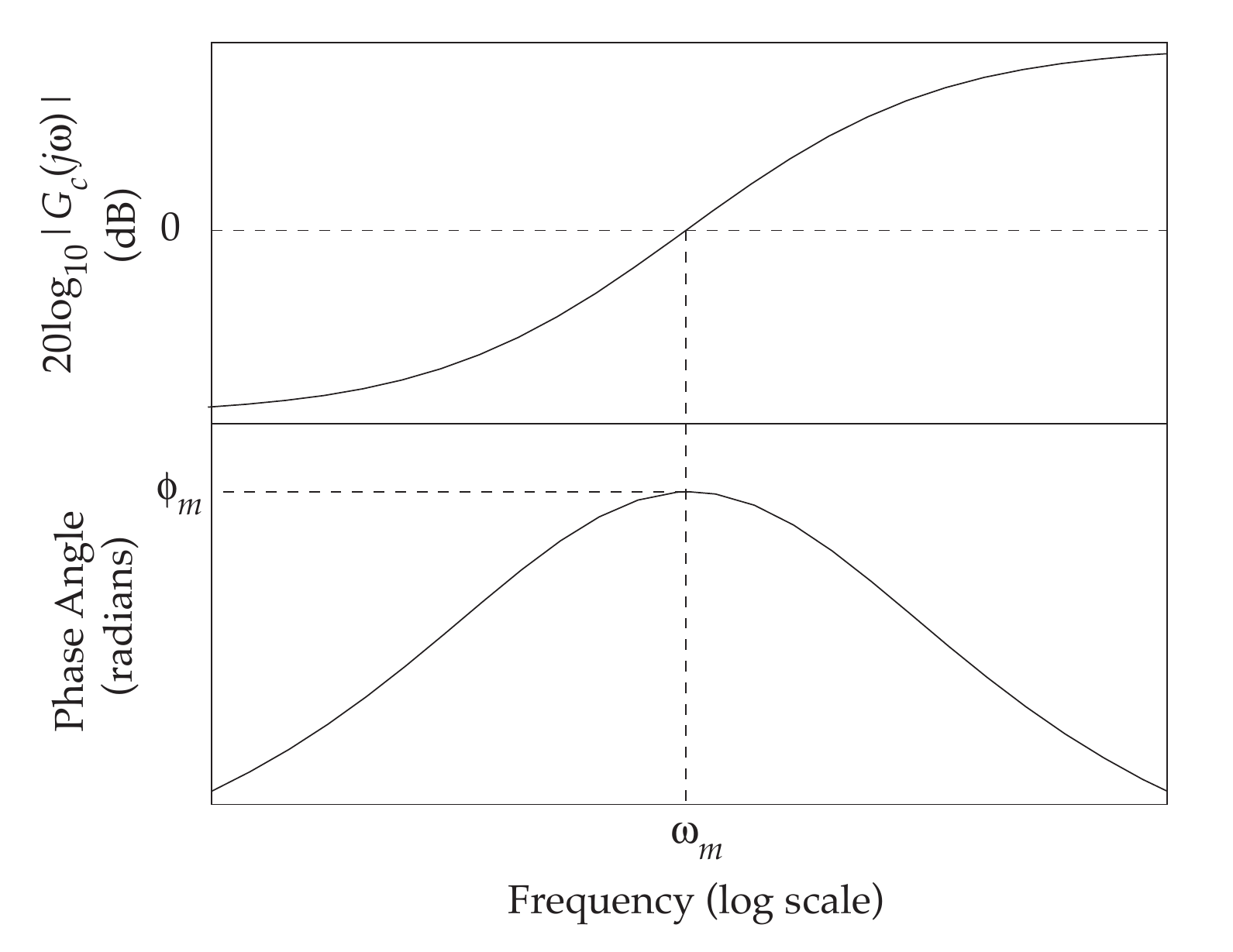} 
	\caption{Phase-lead compensation Bode diagram} 
	\label{Phase Comp Picture} 
	\end{figure}

As shown in Fig.\ \ref{Phase Comp Picture}, at the center frequency 
$\omega_{m}$ the magnitude of the transfer function is unity, and the 
phase angle is $\phi_{m}$.  By assigning $\omega_{m}$ to the 
controller natural frequency, and concatenating this compensator with 
the controller definition in Eq.\ (\ref{CLaw}) phase lags of up to 60 
degrees can be corrected.

\subsubsection{Tustin's Approximation}
Having considered all of the components in the control loop, we derive 
a practical discrete-time equivalent of the control law (\ref{CLaw}).  
A variety of transformation methods can be used: numerical 
integration, zero-pole matching, hold equivalents, etc.  Here we will use a 
trapezoidal numerical integration technique, known in control circles 
as Tustin's method or the bilinear transformation.

Given a continuous-time transfer function $H(s)$, the $z$-transform of 
a Tustin discrete equivalent can be found by substitution
	\begin{equation} 
	H(z)=H(s) \left | \begin{array}{l}
	 \\
	 s={2 \over T}{z-1 \over z+1}
	 \end{array} \right.
	 \label{tustin}
	\end{equation} 
where $T$ is the sampling interval.

The resulting discrete transfer function $H(z)$ is the ratio of two 
polynomials in $z$ and is of the form
	\begin{equation} 
	H(z)={\displaystyle \sum_{k=0}^{ M} b_{k}z^{-k} \over 
	  \displaystyle \sum_{k=0}^{N}a_{k}z^{-k}}
	\end{equation} 

For purposes of controller computation, the constants $a_{k}$ and $b_{k}$ are 
arranged into the corresponding difference equation of the form shown in 
(\ref{DiffEq}).
 
\subsubsection{Prewarping}
The bilinear transformation maps the stable region of the $s$-plane 
(the left half plane) exactly into the finite region within the unit 
circle of the $z$-plane.  Clearly, distortion takes place in this 
mapping, such that the frequency responses of $H(s)$ and $H(z)$ 
differ.  We can force the discrete equivalent to match $H(s)$ at the 
controller natural frequency by ``prewarping'' the continuous-time 
transfer function.  Expressing $H(s)$ as $H({s \over \omega_{oc}})$, 
replace $\omega_{oc}$ with $\omega_{a}$ such that
	\begin{equation} 
	\omega_{a}={2 \over T}\tan{{\omega_{oc}T \over 2}}
	\label{warp}
	\end{equation}
Notice that $\omega_{a}$ approaches $\omega_{oc}$ for high sampling rates.  
Stated as a frequency substitution the prewarped Tustin equivalent is 
described as
	\begin{equation} 
	H(z)=H({s \over \omega_{oc}}) \left | \begin{array}{l}
	 \\
	  {s \over \omega_{oc}}={1 \over \tan{(\omega_{oc}T/2})}{z-1 \over z+1}
	 \end{array} \right.
	 \label{TustinWarp}
	\end{equation}

\subsection{The Complete  Controller}\label{}
Finally, the complete discrete-time controller can be assembled.  To 
maintain the loop gain $G(s)H(s)$ in Fig.\ \ref{SNR Picture}, we must 
adjust the original control law (\ref{CLaw}) to account for the 
additional gain elements in Fig.\ \ref{Implementation Fig}.  Combining 
the optimal controller with the phase-lead compensator we have the 
continuous-time control element.
	\begin{eqnarray}
	\begin{array}{ll}
	H(s)G_{c}(s) & ={\displaystyle {K_{oc} \over K_{a}K_{pa}K_{c}K_{i}\sqrt{\eta}} } 
	\times \vspace{.1in}\\	
	& {\displaystyle 
	 {(s+z_{oc})(1+\eta\tau s) \over 
	(s^{2}+(\omega_{oc}/Q_{oc})s+\omega_{oc}^{2})(1+\tau s)}
	}
	\end{array}
	\end{eqnarray}

The  coefficients of the corresponding difference equation (\ref{DiffEq}) are 
derived using the prewarped Tustin's approximation (\ref{TustinWarp})
	\begin{eqnarray}
	a_{0} & = & 1\\
	\nonumber a_{1} & = & 
	{
	(-\omega^{\prime}-3\tau^{\prime}+3\omega^{\prime 3}+\omega^{\prime 
	2}\tau^{\prime})Q_{oc}+\omega^{\prime 2}-\omega^{\prime}\tau^{\prime}
	 \over 
	 (\omega^{\prime 3}+\omega^{\prime 2}\tau^{\prime}+\omega^{\prime 
	 }+\tau^{\prime})Q_{oc}+\omega^{\prime 2}+\omega^{\prime}\tau^{\prime}
	 }\\
	\nonumber a_{2} & = & 
	{
	(-\omega^{\prime}+3\tau^{\prime}+3\omega^{\prime 3}-\omega^{\prime 
	2}\tau^{\prime})Q_{oc}-\omega^{\prime 2}-\omega^{\prime}\tau^{\prime}
	 \over 
	 (\omega^{\prime 3}+\omega^{\prime 2}\tau^{\prime}+\omega^{\prime 
	 }+\tau^{\prime})Q_{oc}+\omega^{\prime 2}+\omega^{\prime}\tau^{\prime}	 }\\
	\nonumber a_{3} & = & 
	{
	(\omega^{\prime}-\tau^{\prime}+\omega^{\prime 3}-\omega^{\prime 
	2}\tau^{\prime})Q_{oc}-\omega^{\prime 2}+\omega^{\prime}\tau^{\prime}
	 \over 
	 (\omega^{\prime 3}+\omega^{\prime 2}\tau^{\prime}+\omega^{\prime 
	 }+\tau^{\prime})Q_{oc}+\omega^{\prime 2}+\omega^{\prime}\tau^{\prime}	 }
        \label{a-coeffs}
	\end{eqnarray}
and
	\begin{eqnarray}
	b_{0} & = & {Q_{oc}\omega^{\prime}K^{\prime} \over 
	K_{a}K_{pa}K_{c}K_{i}\sqrt{\eta} } \times \\
	&& \nonumber {
	\omega^{\prime}+\eta\tau^{\prime}+\omega^{\prime}\eta 
	z_{oc}^{\prime}\tau^{\prime}+\omega^{\prime 2}z_{oc}^{\prime}
	\over
	 (\omega^{\prime 3}+\omega^{\prime 2}\tau^{\prime}+\omega^{\prime 
	 }+\tau^{\prime})Q_{oc}+\omega^{\prime 2}+\omega^{\prime}\tau^{\prime}	 }
	\\
	\nonumber b_{1} & = & b_{0}
	{
	\omega^{\prime}-\eta\tau^{\prime}+\omega^{\prime}\eta 
	z_{oc}^{\prime}\tau^{\prime}+3\omega^{\prime 2}z_{oc}^{\prime}
	\over
	 \omega^{\prime}+\eta\tau^{\prime}+\omega^{\prime}\eta 
	 z_{oc}^{\prime}\tau^{\prime}+\omega^{\prime 2} z_{oc}^{\prime} }
	\\
	\nonumber b_{2} & = & -b_{0}
	{
	\omega^{\prime}+\eta\tau^{\prime}+\omega^{\prime}\eta 
	z_{oc}^{\prime}\tau^{\prime}-3\omega^{\prime 2}z_{oc}^{\prime}
	\over
	 \omega^{\prime}+\eta\tau^{\prime}+\omega^{\prime}\eta 
	 z_{oc}^{\prime}\tau^{\prime}+\omega^{\prime 2} z_{oc}^{\prime} }
	\\
	\nonumber b_{3} & = & -b_{0}
	{
	\omega^{\prime}-\eta\tau^{\prime}+\omega^{\prime}\eta 
	z_{oc}^{\prime}\tau^{\prime}-\omega^{\prime 2}z_{oc}^{\prime}
	\over
	 \omega^{\prime}+\eta\tau^{\prime}+\omega^{\prime}\eta 
	 z_{oc}^{\prime}\tau^{\prime}+\omega^{\prime 2}	z_{oc}^{\prime} }
        \label{b-coeffs}
	\end{eqnarray}
where the normalized parameters are: \newline 
      $K^{\prime}={K_{oc} \over \omega_{oc}}$, 
      $\tau^{\prime}={1 \over \sqrt{\eta}}$, 
	  $z_{oc}^{\prime}={z_{oc} \over \omega_{oc}}$, and 
	  $\omega^{\prime}=\tan{{\omega_{oc}T \over 2}}$.

A final note: Occasionally the controller output $y(n)$, calculated 
from the difference equation, will exceed the D/A converter range 
$[0, C_{da}]$.  Since some converters wrap overranged values 
to the opposite edge of the range, it is essential that the output 
value be limited:
	\begin{equation}
	y(n)_{out}=\max{(\min{(y(n),C_{da})},0)}
	\end{equation}
	

\end{document}